\pdfoutput=1
\documentclass[11pt]{article}
\usepackage[utf8]{inputenc}
\usepackage[T1]{fontenc}
\usepackage{fixltx2e}
\usepackage{graphicx}
\usepackage{longtable}
\usepackage{float}
\usepackage{wrapfig}
\usepackage{rotating}
\usepackage[normalem]{ulem}
\usepackage{amsmath}
\usepackage{textcomp}
\usepackage{marvosym}
\usepackage{wasysym}
\usepackage{amssymb}
\usepackage{hyperref}
\usepackage{microtype}
\usepackage[utf8]{inputenc}
\usepackage[T1]{fontenc}
\usepackage{subcaption}
\graphicspath{{figures/},{figures-old/}}
\usepackage{fancyhdr}
\makeatletter
\let\ps@plain\ps@fancy
\makeatother

\author{Brian L. DeCost, Toby Francis, Elizabeth A. Holm}
\date{\today}
\title{Exploring the microstructure manifold: image texture representations applied to ultrahigh carbon steel microstructures}
\hypersetup{
  pdfkeywords={},
  pdfsubject={},
  pdfcreator={Emacs 24.5.1 (Org mode 8.2.10)}}
\begin{document}

\maketitle

\begin{abstract}
We introduce a microstructure informatics dataset focusing on complex, hierarchical structures found in a single Ultrahigh carbon steel under a range of heat treatments.
Applying image representations from contemporary computer vision research to these microstructures, we discuss how both supervised and unsupervised machine learning techniques can be used to yield insight into microstructural trends and their relationship to processing conditions.
We evaluate and compare keypoint-based and convolutional neural network representations by classifying microstructures according to their primary microconstituent, and by classifying a subset of the microstructures according to the annealing conditions that generated them.
Using t-SNE, a nonlinear dimensionality reduction and visualization technique, we demonstrate graphical methods of exploring microstructure and processing datasets, and for understanding and interpreting high-dimensional microstructure representations.
\end{abstract}
\section{Introduction}
\label{sec-1}
Comprised of the structures that arise from processing and mediate properties, microstructure is a core focus of the discipline of materials science.
Microstructural information is most often conveyed via images obtained through various microscopy techniques (i.e. micrographs), sometimes supplemented by other structural and compositional probes.
Traditionally, microstructural images have been evaluated by human experts, both to interpret the micrographs themselves and to connect them to processing conditions and property outcomes.
However, recent research in microstructure informatics has begun to explore applications of contemporary computer vision to construct microstructure representations suitable for use in machine learning and microstructure analytics tasks\cite{decost2015,decost2017jom,chowdhury2016,lubbers2016}.
For example, \cite{chowdhury2016} compare several image texture representations and find that off-the-shelf convolutional neural network (CNN) features can be applied to microstructure analytics tasks (e.g. classification) without fine-tuning any of the CNN parameters.
Likewise, Lubbers et al.\cite{lubbers2016} apply bilinear CNN representations\cite{gatys2015,lin2015bilinear,lin2015} to synthetic lamellar structures, and relate this representation to the generative microstructure model parameters (i.e. lamellar spacing and orientation and noise).
While promising proofs of principle, these studies used comparatively simple and well-parameterized microstructures.
To move towards quantitative application of generic computer vision techniques, we require real-world, technologically-relevant microstructure systems exhibiting the complex, hierarchical structures that challenge conventional microstructure segmentation and quantification.

To this end, we introduce the CMU-UHCS (Carnegie Mellon University Ultrahigh Carbon Steel) dataset\footnote{to appear on \url{https://materialsdata.nist.gov}}, based on the work of Hecht et al.\cite{hecht2016,hecht2017}.
This dataset consists of 961 scanning electron microscopy (SEM) micrographs of Ultrahigh Carbon Steel (UHCS) subjected to a variety of heat treatments and taken at several different magnifications.
The dataset spans several complex and hierarchical microconstituents typically found in UHCS and other technologically relevant alloy systems, offering a compelling real-world microstructure informatics challenge.

UHCS (steels with 1-2.1 wt\% carbon) are intermediate in content to high carbon steel (0.6-1 wt\% C) and cast iron (2.1-4.3 wt\% C).
Due to their high carbon content relative to conventional steels, a characteristic microstructure feature of these alloys is proeutectoid cementite (Fe$_{\text{3}}$C), typically forming a carbide network associated with the grain boundaries of the high-temperature austenite phase.
The hard, brittle carbides help lend UHCS its well-known high strength and wear resistance, but highly-connected intergranular carbide networks can be detrimental to toughness and ductility by providing extended pathways for crack propagation\cite{sherby1999,hyzak1976}.
Recent UHCS research has focused on mitigating this weakness by optimizing the network microstructure through various heat treatments\cite{pacyna2007} and addition of minor alloying elements\cite{wang2007,liu2011}.
Hecht et al. recently developed a quantitative measure of the carbide network connectivity, relating this to annealing schedules and toughness measurements\cite{hecht2016}.
A similar study concerning the effect of annealing conditions on spheroidite morphology is forthcoming\cite{hecht2017}.
The present UHCS microstructure dataset is built on the characterization  efforts for these two UHCS studies.

In this study, we use the UHCS dataset compare state-of-the-art CNN-based image texture representations with the classic bag of visual words (BoW) representation\cite{csurka2004,zhang2007}.
As microstructure representations, the BoW has the theoretical advantage of strong explicit scale and rotation invariance, while CNNs notoriously outperform BoW on typical natural image recognition tasks (e.g. facial recognition, object detection and identification, scene classification).
We evaluate each image representation using both supervised and unsupervised learning methods, and demonstrate how these techniques can be used together for exploratory microstructure analysis.
Specifically, we used a Support Vector Machine (SVM)\cite{cortes1995} approach to classify microstructures both by primary microconstituent and annealing condition.
We complement this understanding by applying the unsupervised dimensionality-reduction technique t-SNE (t-distributed Stochastic Neighbor Embedding)\cite{vandermaaten2008} to visualize the high-dimensional distributions of each microstructure representation, relating this structure to available annealing schedule and imaging metadata.

Our primary contributions in this report are:

\begin{itemize}
\item A real-world dataset of complex, hierarchical microstructures annotated with  microstructure constituent metadata, as well as partial imaging and processing metadata, such as heat treatment, quenching procedure, and magnification.
\item Evaluation of several competitive computer vision techniques, with discussion of their relative strengths and weaknesses for a range of real-world microstructure informatics tasks.
\item Exploration of these microstructure representations for inferring processing -- microstructure -- properties relationships for realistic complex, hierarchical microstructure systems.
\end{itemize}

\section{Methods}
\label{sec-2}
\subsection{UHCS Dataset}
\label{sec-2-1}
\begin{figure*}[!htbp]
  \centering
  \begin{subfigure}[]{0.32\textwidth}
  \includegraphics[width=\textwidth]{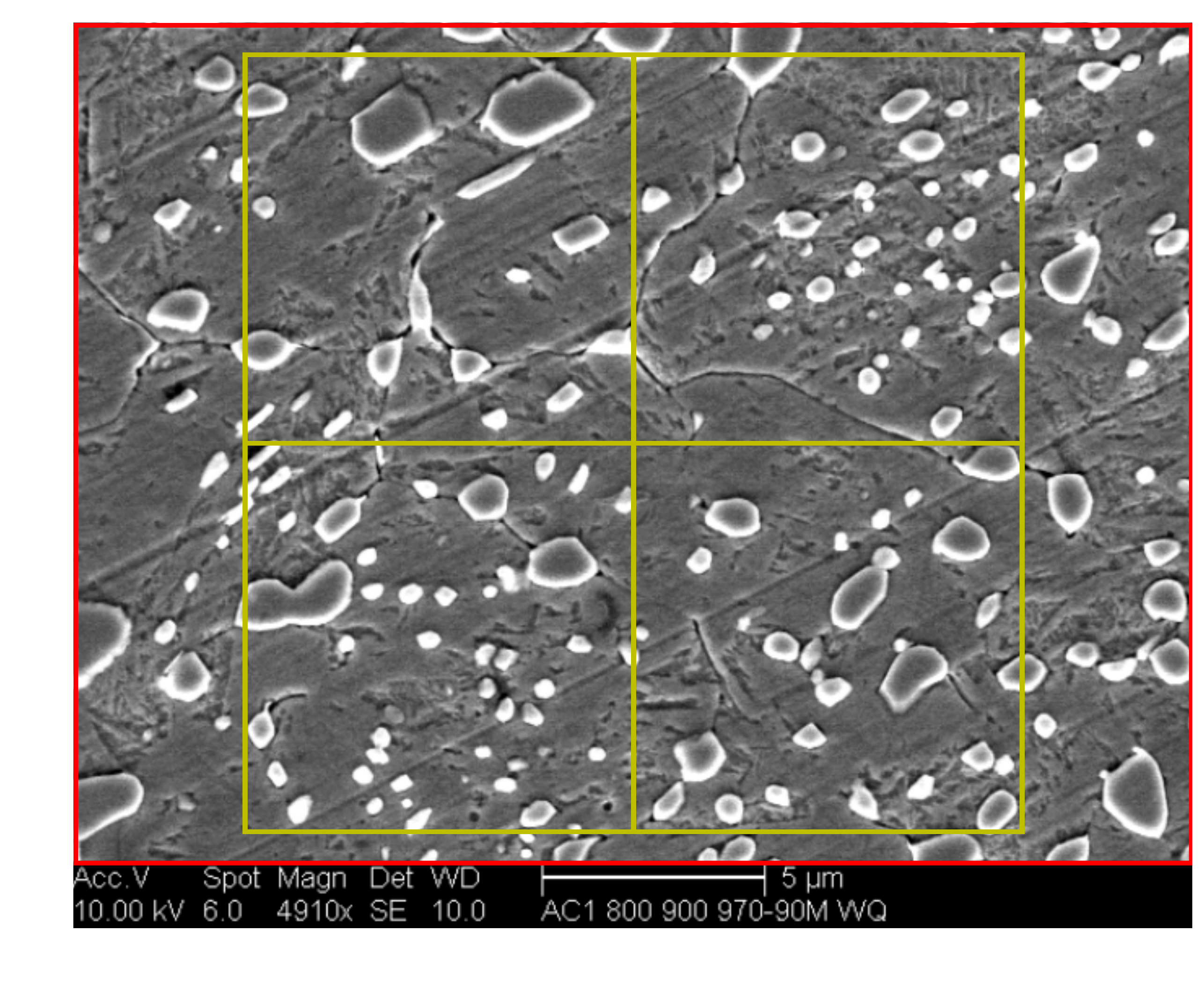}
  \caption{}
  \label{fig:boundarycementite}
  \end{subfigure}
  \begin{subfigure}[]{0.32\textwidth}
  \includegraphics[width=\textwidth]{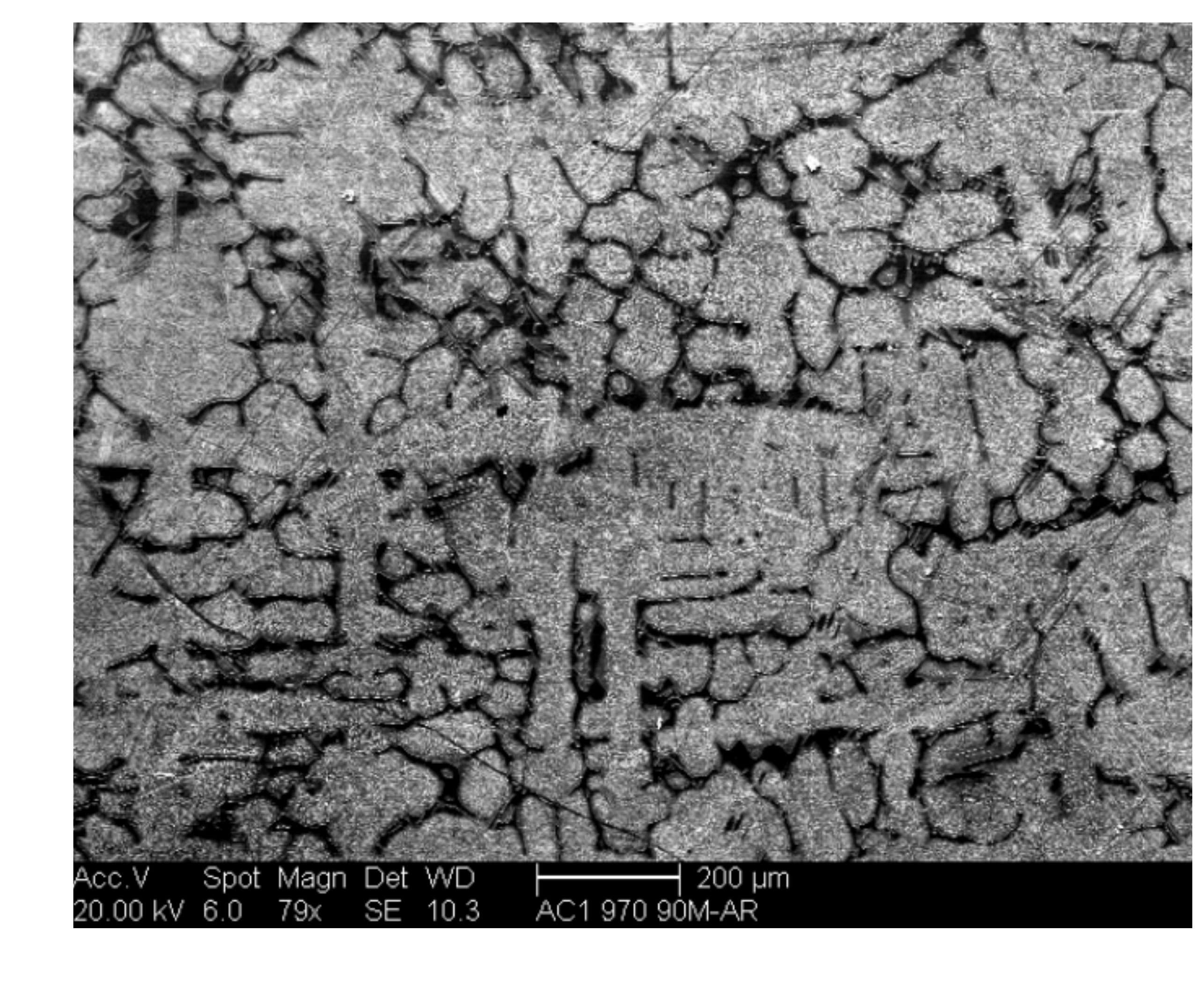}
  \caption{}
  \label{fig:carbidenetwork}
  \end{subfigure}
  \begin{subfigure}[]{0.32\textwidth}
  \includegraphics[width=\textwidth]{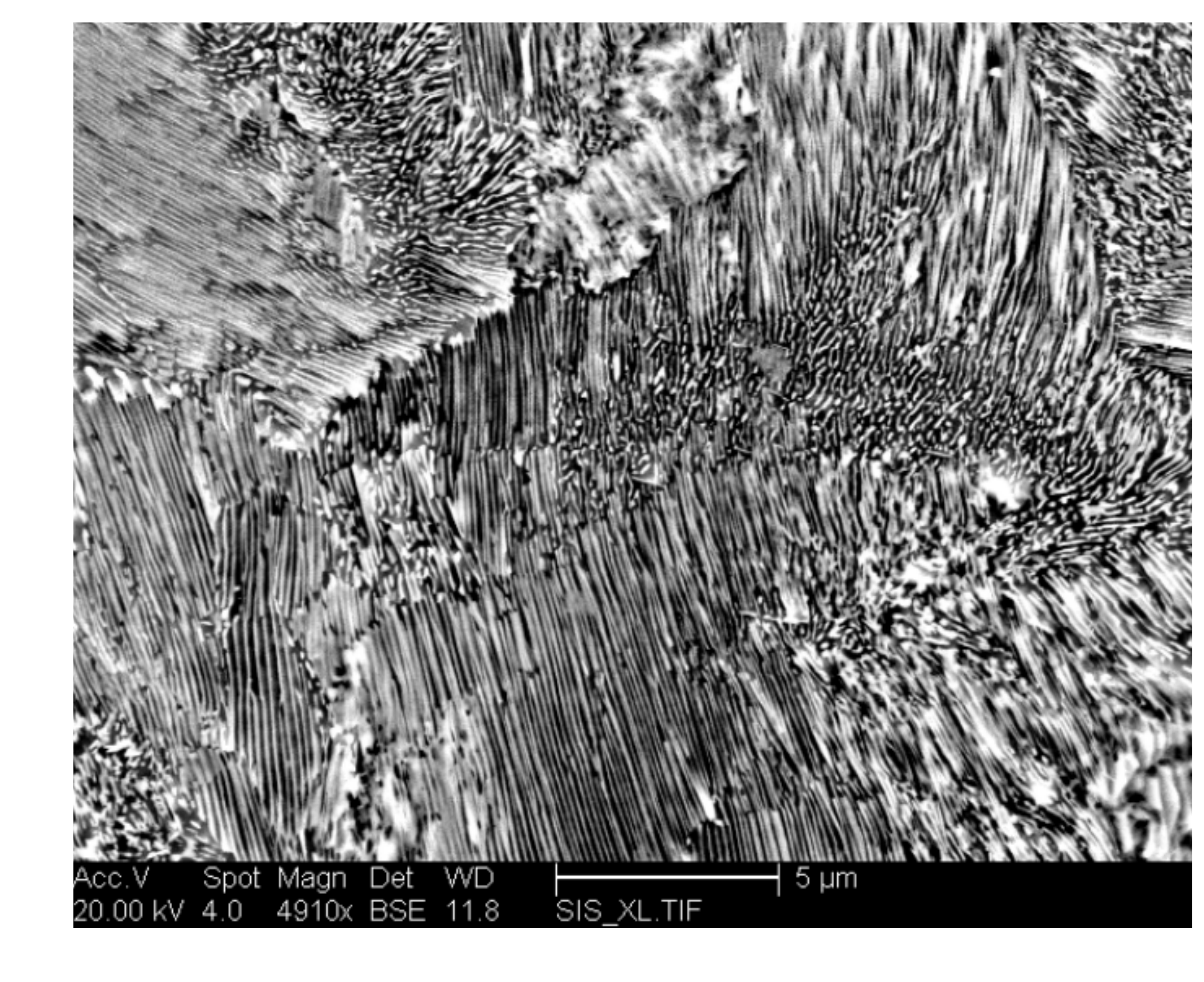}
  \caption{}
  \label{fig:pearlite}
  \end{subfigure} \\
  \begin{subfigure}[]{0.32\textwidth}
  \includegraphics[width=\textwidth]{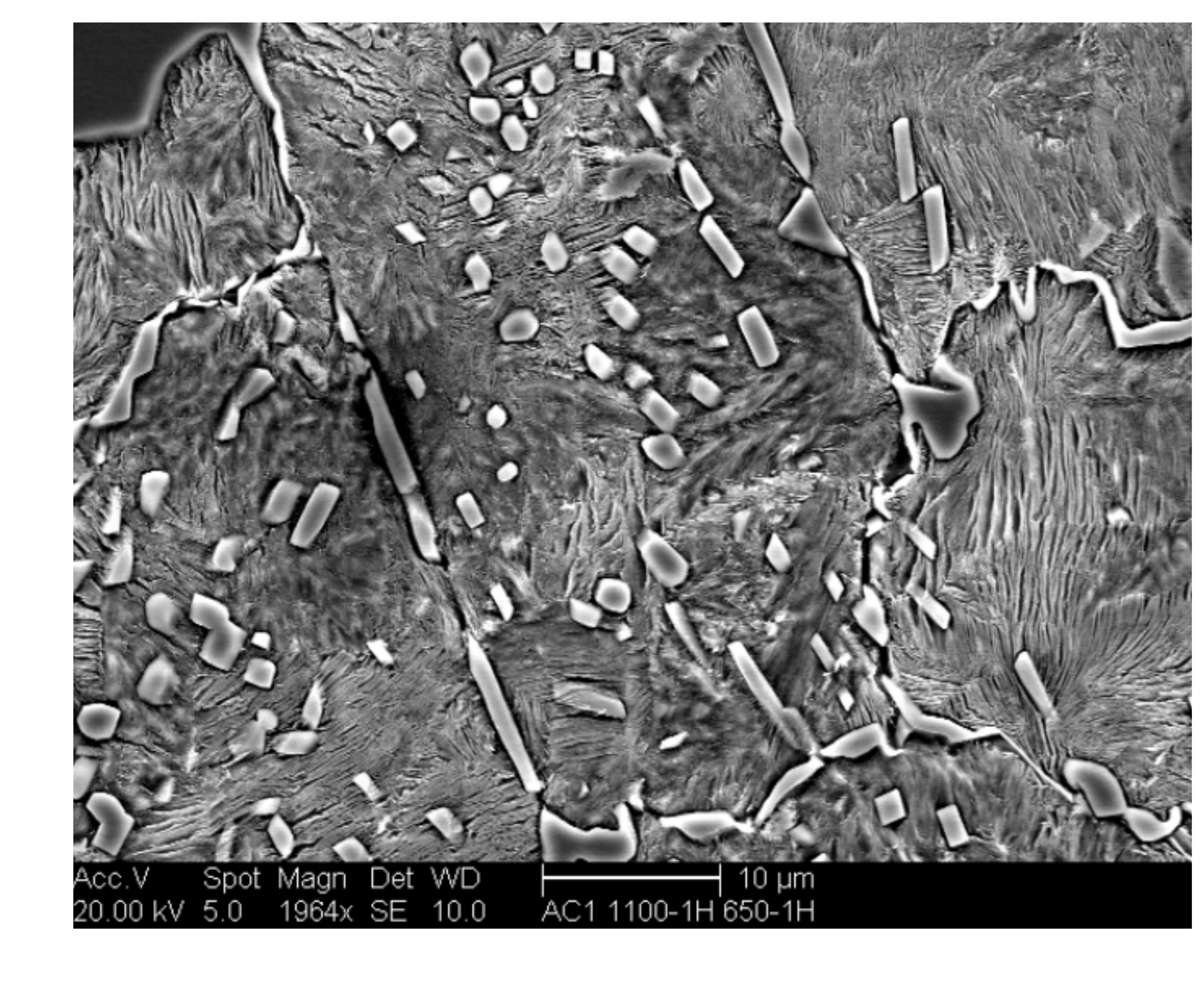}
  \caption{}
  \label{fig:pearlite+cementite}
  \end{subfigure}
  \begin{subfigure}[]{0.32\textwidth}
  \includegraphics[width=\textwidth]{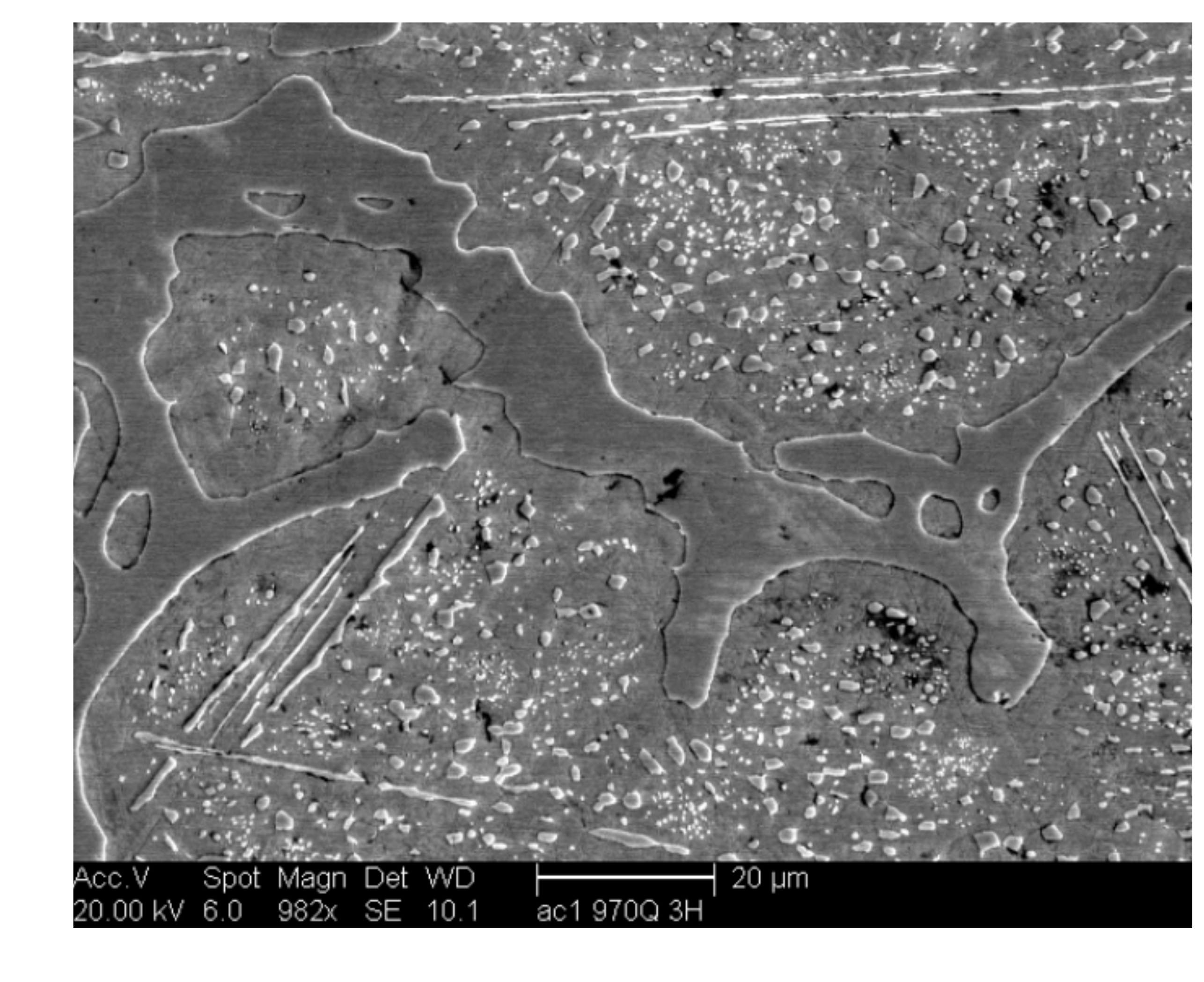}
  \caption{}
  \label{fig:cementite+widmanstatten}
  \end{subfigure}
  \begin{subfigure}[]{0.32\textwidth}
  \includegraphics[width=\textwidth]{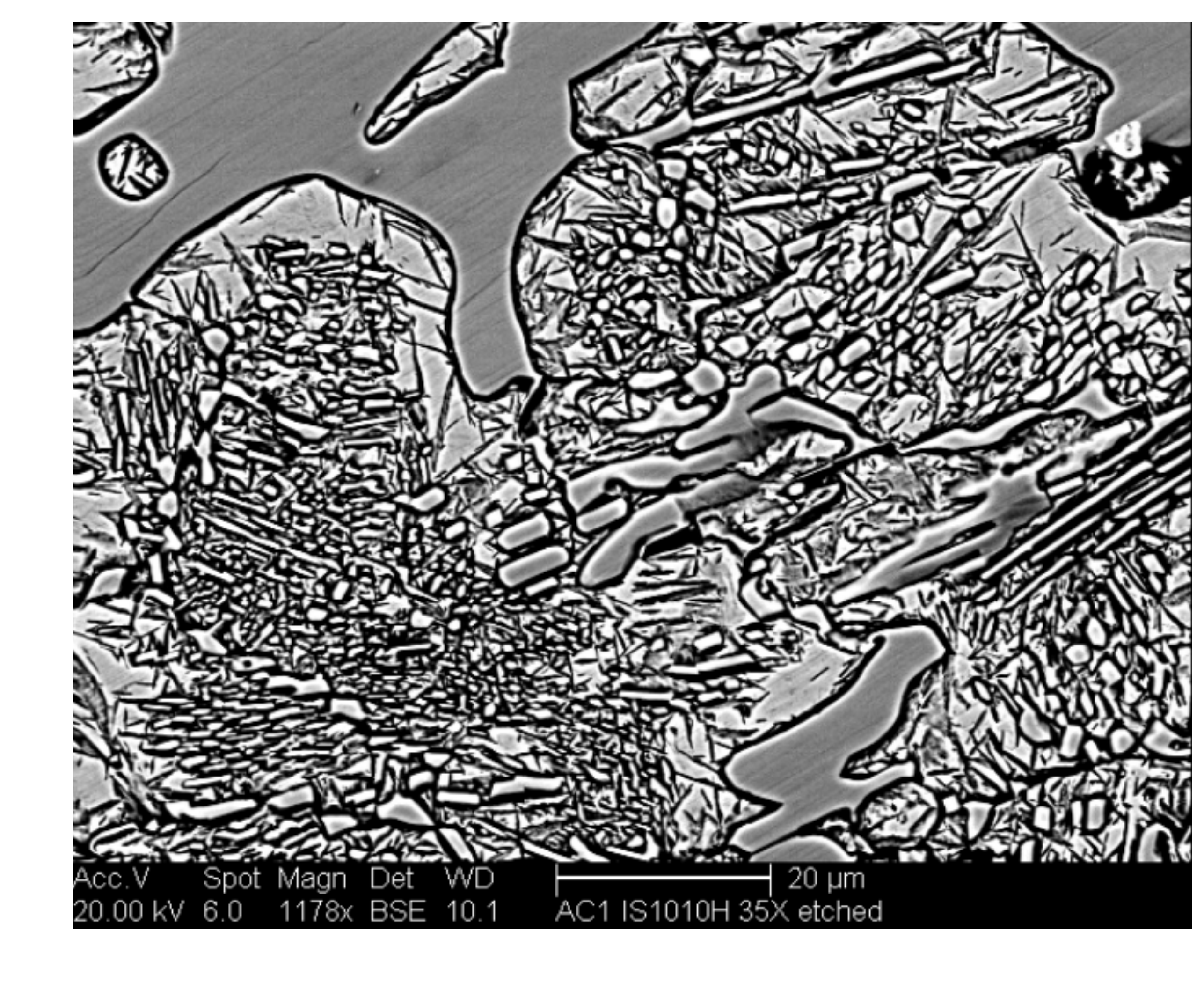}
  \caption{}
  \label{fig:martensite}
  \end{subfigure}
  \caption{Primary microstructure constituents in the UHCS dataset: (\subref{fig:boundarycementite}) spheroidized cementite with red and yellow frames indicating image regions used for feature extraction in the UHCS-600 and UHCS-2400 datasets, (\subref{fig:carbidenetwork}) carbide network microstructure, (\subref{fig:pearlite}) pearlite, (\subref{fig:pearlite+cementite}) pearlite containing spheroidized cementite, (\subref{fig:cementite+widmanstatten}) Widmanstätten cementite, and (\subref{fig:martensite}) martensite and/or bainite.}
  \label{fig:representativemicrographs}
\end{figure*}

The UHCS dataset consists of 961 SEM micrographs of commercial UHCS subjected to a range of heat treatments by Hecht et al.\cite{hecht2016,hecht2017}.
These micrographs span a wide range of magnifications, and include both secondary electron (SE) and back-scattered electron (BSE) images.
598 micrographs also have annealing schedule metadata: annealing time, temperature, and quench medium.
All 961 images are labeled with their primary microstructure constituents as illustrated in Figure \ref{fig:representativemicrographs}.
Most of the micrographs focus on the spheroidite morphology (Figure \ref{fig:boundarycementite}), the carbide network (Figure \ref{fig:carbidenetwork}), and pearlite (Figure \ref{fig:pearlite}).
A smaller number of micrographs contain two primary microconstituents, such as pearlite containing spheroidite (Figure \ref{fig:pearlite+cementite}), Widmanstätten cementite (Figure \ref{fig:cementite+widmanstatten}), and martensite (Figure \ref{fig:martensite}).
Table \ref{tab:dataset} shows the distribution of each of these primary microconstituent labels.
We used the full set of 961 labeled $645 \times 484$ pixel micrographs to generate the data visualizations in Section \ref{sec:datavis}.

\begin{table}[htb]
\caption{\label{tab:dataset} Schedule of primary microconstituent labels in the UHCS micrograph dataset.}
\centering
\begin{tabular}{lc}
primary microconstituents & \# of micrographs \\
\hline
spheroidite & 374 \\
carbide network & 212 \\
pearlite & 124 \\
pearlite + spheroidite & 107 \\
Widmanstätten cementite & 81 \\
pearlite + Widmanstätten & 27 \\
Martensite/Bainite & 36 
\end{tabular}
\end{table}

For the primary microconstituent classification experiments, we considered only a subset of these labeled micrographs: 200 randomly selected micrographs each from the spheroidized cementite, carbide network, and pearlite/pearlite+spheroidite classes, for a total of 600 images.
We also consider an expanded dataset constructed by cropping four $224 \times 224$ sub-images from the center of each micrograph in the original dataset, so that the expanded dataset consists of 2400 images.
The single red and four yellow frames in Figure \ref{fig:boundarycementite} indicate the image regions used for microstructure feature extraction in the full-sized and cropped image sets, respectively.

The annealing schedule classification task was limited to the micrographs collected to study the spheroidite morphology.
The dataset contains spheroidite micrographs resulting from 23 distinct annealing schedules.
Within this subset of micrographs, we limit the classification dataset to the 13 annealing conditions with at least 15 micrographs.
Where more than 15 micrographs with a given annealing condition are available, we randomly select 15 micrographs to obtain a balanced classification dataset.
The resulting annealing condition classification datasets consist of 195 full-sized micrographs and 780 cropped micrographs.

\subsection{Image representations \label{sec:featureextraction}}
\label{sec-2-2}
In this work, we explore and compare two computer vision approaches for computing generic image representations: Mid-level image patch descriptors\cite{zhang2007,li2008} and convolutional neural network (CNN) representations\cite{lecun2015, guo2015, schmidhuber2015}.
The mid-level features approach is attractive due to its relatively strong invariance to image scale and orientation; its focus on identifying and characterizing individual features is also intuitive to the materials scientist.
However, CNN representations are generally regarded as richer, more hierarchical, and more effective than mid-level image features, even when transferring CNN parameterizations from one task to another (in this case completely unrelated) task.

\subsubsection{Mid-level image features \label{sec:bow}}
\label{sec-2-2-1}
The baseline feature extraction method in this study is the bag of visual words (BoW) method, which represents an image as a distribution of local image descriptors (i.e. visual features).
As previously reported in detail\cite{decost2015,decost2017jom}, we applied both the Difference of Gaussians\cite{lowe1999} and the Harris-LaPlace\cite{mikolajczyk2004} interest point detectors to select distinctive image regions with characteristic scales and orientations.
We then used oriented SIFT descriptors\cite{lowe1999} to characterize the visual appearance of each interest point, and k-means clustering\cite{lloyd1982} to quantize the SIFT descriptors into a visual dictionary with 100 (BoW$_{\text{100}}$) visual words (i.e. SIFT cluster centers).
Each image is then represented by its microstructural fingerprint: A normalized histogram measuring the occurrence frequency of each visual word within the image.
BoW representations can be compared with various similarity metrics for discrete probability distributions, such as the Hellinger and $\chi^2$ kernels\cite{zhang2007}.
Presently, we use the $\chi^2$ kernel, which for two normalized $m$-dimensional histograms $X = \{x_i\}$ and $Y = \{y_i\}$ is written as:

\begin{equation}
  D_{\chi^2} \left( X, Y \right) = \frac{1}{2} \sum_{i=1}^{m} \frac{(x_i - y_i)^2}{x_i + y_i}
\end{equation}

Typically the exponential version of the $\chi^2$ is used with SVM classification:
\begin{equation}
K_{\chi^2}(X,Y) = \exp\left(-\frac{1}{A} D_{\chi^2}\left(X,Y\right)\right)
\end{equation}

We follow \cite{zhang2007} by setting the kernel parameter $A$ to the mean $\chi^2$ distance between BoW representations of all of the training examples.

In addition to sparse, oriented SIFT features computed at interest points (s-SIFT), we also used dense sampling of SIFT features (d-SIFT) at multiple scales, with fixed orientation.
This results in a much larger set of local features for constructing BoW representations and lends itself to more efficient numerical implementation\cite{jurie2005}.
In some applications, this can lead to improved recognition performance, even though the individual region descriptors are no longer rotation invariant.

Ambiguity between visual words is a major weakness of the BoW methods: in general, boundaries between clusters in the visual dictionary are arbitrary\cite{boureau2010} in the sense that they are a convenient means of discretizing a high-dimensional vector space rather than based on a physically meaningful set of visual features\cite{vangemert2010}.
The result is that a feature located near a cluster boundary may be very far from representative of that cluster of features, yet it is weighted the same as a feature near the cluster centroid.
Worse still, a very similar feature that falls on the other side of the cluster boundary will be assigned to a different visual word!
There are many techniques that attempt to mitigate the resulting visual word assignment uncertainty, for example by assigning local feature descriptors to multiple visual words\cite{philbin2008} or by using a probabilistic visual dictionary\cite{boureau2010,koniusz2013}.
Other methods record more detailed information about the relationships between the visual dictionary and local feature descriptors extracted from an image\cite{boureau2010pooling}.
In this report, we explore one simple and effective method: VLAD (Vector of Locally Aggregated Descriptors)\cite{jegou2010}.

\subsubsection{VLAD encoding \label{sec:vlad}}
\label{sec-2-2-2}
The Vector of Locally Aggregated Descriptors (VLAD)\cite{jegou2010} technique is closely related to the BoW method.
VLAD attempts to mitigate the ambiguity between visual words by recording the \emph{difference} between local feature descriptors and the corresponding visual word (i.e. the cluster centroid), rather than simply constructing a distribution of visual word occurrence frequency.
After assigning all the local feature descriptors to visual words, VLAD sums up the residual vectors (the differences) between each visual word and all the local features assigned to it.
The final VLAD descriptor is obtained by concatenating the residual sums for each visual word.
We applied the block-wise normalization scheme (intra-normalization) from \cite{arandjelovic2013}: each residual sum is L2-normalized before being concatenated, and the resulting VLAD vector is L2-normalized as well.

We used VLAD to encode both sparse and dense SIFT features extracted using the same methods outlined in Section \ref{sec:bow}.
With VLAD encoding, it is common to use a smaller visual dictionary size; here we used dictionary sizes of 32 (VLAD$_{\text{32}}$) and 64 (VLAD$_{\text{64}}$).
This results in VLAD features with $128\times32 = 4096$ and $128\times64 = 8192$ dimensions, respectively.

\subsubsection{Convolutional Neural Network features}
\label{sec-2-2-3}
In the past few years, convolutional neural networks (CNN) have demonstrated excellent performance at many computer vision tasks; this success is often credited to the hierarchical nature of the image representation they construct.
CNNs extract high-level image features by stacking multiple layers of neurons organized into convolution filters learned from annotated training images.
By interleaving pooling (effectively down-sampling) steps between layers of convolution filters, CNNs obtain hierarchical representations of image content.

Though CNNs are notorious for requiring extreme amounts of training data (and for overfitting on small datasets), recent research efforts in \emph{transfer learning}\cite{pan2010} have shown that deep CNNs can generalize well to new datasets, in some cases even when new task is not related to the original task\cite{donahue2013}.
Simple approaches include using the output of the high-level CNN layers as input to a linear SVM\cite{razavian2014}, or retraining (fine-tuning) some or all of the layers of the pre-trained CNN using a new training set.
In this study we use high-level features the from the VGG16 CNN architecture\cite{simonyan2014}, parameterized for object recognition on the ImageNet ILSVRC-2014 dataset\cite{russakovsky2015}, which consists of approximately 1.2 million images representing 1000 object categories (none of which include microstructures).
The VGG16 architecture consists of 14 convolution layers arranged into 5 blocks delineated by pooling (upsampling) layers, followed by two fully-connected layers of 4096 neurons each, and a final 1000-class classification layer.
Because the VGG16 CNN operates on color images, we preprocess each SEM micrograph by replicating the raw grayscale image in each color channel of a new RGB image and subtracting the average intensity of the ImageNet training set for each channel, as recommended by \cite{simonyan2014}.
We used the publicly available parameters provided by the VGG group\cite{simonyan2014} without any fine-tuning.

Fully-connected CNN features were not computed for the full-sized images, because the fully-connected layers fix the size of allowable input images to the size of the training set images.
This can be mitigated by pooling the fully-connected CNN features from appropriately-sized regions within a larger image\cite{gong2014}.
However, for transfer learning tasks (and particularly for image texture recognition) it is much more efficient and effective to apply pooling to the high-level convolution layers\cite{cimpoi2015}, which can be easily extracted from input images of arbitrary size.

We investigated the third convolution layer from both the fourth (VGG$_{\text{4}}$) and fifth (VGG$_{\text{5}}$) convolution blocks of the VGG16 architecture.
For this neural network, both the VGG$_{\text{4}}$ and VGG$_{\text{5}}$ convolution blocks produce 512-channel feature maps, respectively sized $14 \times 14$ and $7 \times 7$ for the cropped UHCS input images and $40 \times 30$ and $20 \times 15$ for the large UHCS input images.
We used VLAD encoding with both a 32-element (VLAD$_{\text{32}}$) and a 64-element (VLAD$_{\text{64}}$) dictionary on the VGG$_{\text{4}}$ and VGG$_{\text{5}}$ feature maps, yielding VLAD vectors of length $512\times32 = 16384$ and $512\times64 = 32768$ respectively.
One advantage of using an encoding method such as VLAD on convolution features is that it yields a deep representation of the image structure with no explicit high-level spatial dependence -- a desirable property for image texture (and microstructure) recognition\cite{cimpoi2015}.

Finally, we explore a simple technique to increase the scale invariance of these CNN representations.
For each input image, we apply VLAD encoding to VGG$_{\text{4}}$ and VGG$_{\text{5}}$ feature maps from four scales, yielding multiscale CNN representations (mVGG$_{\text{4}}$ and mVGG$_{\text{5}}$).
We use bilinear interpolation to downsample the original resolution twice by a factor of $\sqrt{2}$, and to upsample the original resolution once by the same factor.
A greater degree of scale-invariance can be achieved by pooling over a finer-grained set of image scales.

\subsection{SVM classification}
\label{sec-2-3}
We use Support Vector Machine (SVM) classification for microstructure categorization on the UHCS dataset.
We compared BoW representations using the $\chi^2$ kernel as outlined in Section \ref{sec:bow}.
For each of the other image representations, we used linear SVM classification.
Because SVM classification is sensitive to the absolute scale of the input features, we L2-normalize them and set the SVM margin parameter $C$ to $1$, following\cite{cimpoi2015}.
Reported performance figures are obtained via 10\texttimes{} 10-fold cross-validation on the full dataset; the uncertainties reported are sample standard deviations computed on the 100 validation sets.
The classification results are insensitive to changes in the value of the margin parameter $C$.

\subsection{Data visualization}
\label{sec-2-4}
We visualize each high-dimensional microstructure representation using t-SNE (t-distributed Stochastic Neighbor Embedding)\cite{vandermaaten2008}, a non-parametric visualization technique for high-dimensional data.
t-SNE often better captures high-dimensional structure of real-world data compared with other dimensionality reduction techniques such as principal component analysis (PCA)\cite{jolliffe2002}, multidimensional scaling (MDS)\cite{kruskal1964}, Isomap\cite{tenenbaum2000}, and Locally Linear Embedding\cite{roweis2000}.
Rather than preserving global distances between dissimilar points as in PCA, t-SNE preserves only the local structure and similarity of the data points, using a probabilistic measure of `similarity'.
t-SNE uses the high-dimensional Euclidean distance between data points to define pairwise conditional probabilities for each point being a close neighbor to all others in the dataset.
The similarity of a high-dimensional data point x$_{\text{j}}$ with respect to the data point x$_{\text{i}}$ is modeled as a probability of observing x$_{\text{j}}$ under a gaussian distribution $P_i$ with variance $\sigma_i$ centered at x$_{\text{i}}$: $p_{j|i} \sim \exp\left( -||x_i - x_j||^2  / 2\sigma_i \right)$.
The $\sigma_i$ are chosen so that each gaussian distribution has a fixed perplexity $Perp(P_i) = 2^{H(P_i)} = 2^{-\sum_j p_{j|i}\log_2 p_{j|i}}$, effectively tuning the number of nearest neighbor data points.
The similarity of a map point y$_{\text{i}}$ with respect to another map point y$_{\text{j}}$ is analogous, replacing the gaussian distribution with the student t distribution $Q_i$ with one degree of freedom: $q_{j|i} \sim \left( 1 + ||y_i - y_j||^2 \right)^{-1}$.

t-SNE proceeds by attempting to minimize the mismatch in these conditional neighbor probabilities between the original high-dimensional dataset and the low-dimensional representation.
This is accomplished by minimizing the sum of Kullback-Leibler divergences for each data point: $\min \sum_i KL(P_i||Q_i)  = \sum_i \sum_j p_{j|i} \log \frac{p_{j|i}}{q_{j|i}}$ via gradient descent.
This objective function emphasizes local structure by heavily penalizing large distances between map points where the corresponding high-dimensional distance is small, while effectively ignoring small distances between map points where the high-dimensional distance is large.
Because t-SNE is a stochastic algorithm (unlike PCA), we perform t-SNE ten times for each image representation and select the best map, i.e. the map with the smallest value for the objective function.
We use the same t-SNE map for each representation when drawing maps of class labels and processing parameters.

\section{Results and Discussion}
\label{sec-3}
\subsection{Primary microconstituent classification \label{sec:svmresults}}
\label{sec-3-1}
In order to evaluate various microstructural representations, we classified the UHCS micrographs both by microconstituent categories and by annealing conditions.
The microconstutuent classification was performed separately on the UHCS-600 dataset (600 full-size images, 200 in each of three categories) and the UHCS-2400 dataset (2400 cropped images, 800 in each of three categories).
The annealing schedule classification examined the same datasets, but was limited to the spheroidite images produced by the 13 distinct annealing schedules with at least 15 micrographs.
With larger image sets and fewer classes, the microconstutuent classification is more representative of the attainable accuracy of the methods, while the annealing condition classification challenges the approach in the limit of small datasets.

Table \ref{tab:svm} reports cross-validation accuracies obtained via SVM classification using each of the feature extraction methods outlined in Section \ref{sec:featureextraction}.
The first two data columns show the average validation set accuracies on the UHCS-2400 and UHCS-600 image sets for the primary microconstituent classification task; the second two data columns show the same for the spheroidite annealing condition classification task.
For the microconstituent classification task, accuracies are computed via $10\times$ 10-fold cross-validation, while for the annealing condition classification task accuracies are obtained via a stratified leave-one-out cross-validation scheme.
The uncertainties reported are the standard deviations of the accuracies achieved on the validation and training sets, respectively.
Uncertainties for the annealing schedule classification task are much higher, principally because of the much smaller dataset size.

The classification accuracy of a given feature extraction method is generally consistent between the two UHCS datasets, with a slight (within measurement uncertainty) advantage for the uncropped dataset.
This advantage could potentially be a result of higher variability in the UHCS-2400 dataset, as not all of the original micrographs exhibit homogeneous microstructure features.
For example, the cropped spheroidite images in Figure \ref{fig:boundarycementite} are clearly not representative microstructure samples relative to the full image: the upper left quadrant contains a high proportion of grain boundary cementite, while the upper right quadrant contains almost no grain boundary cementite and a high proportion of interior cementite.

\begin{table*}[htb]
\caption{\label{tab:svm}Quantitative evaluation of microstructure representations. Cross-validation accuracy (\textpm{} standard deviation) for SVM classification. The left two data columns show validation set scores for the primary microconstituent classification task, and the right two columns show the validation set scores for the annealing schedule classification task.}
\centering
\begin{tabular}{lcccc}
 & \multicolumn{2}{c}{microconstituent} & \multicolumn{2}{c}{annealing schedule} \\
method & UHCS-2400 & UHCS-600 & UHCS-2400  & UHCS-600  \\
\hline
raw & 45.0 (\textpm{} 5.36) & 55.3 (\textpm{} 5.73) & 14.5 (\textpm{} 3.07) & 18.5 (\textpm{} 9.99) \\
dSIFT BoW$_{32}$ & 86.9 (\textpm{} 4.65) & 89.0 (\textpm{} 3.82) & 48.7 (\textpm{} 6.53) & 39.5 (\textpm{} 11.2) \\
dSIFT BoW$_{100}$ & 91.3 (\textpm{} 3.76) & 92.8 (\textpm{} 2.87) & 61.0 (\textpm{} 7.32) & 50.8 (\textpm{} 16.1) \\
sSIFT BoW$_{32}$ & 88.8 (\textpm{} 3.53) & 91.2 (\textpm{} 3.26) & 59.9 (\textpm{} 6.2) & 41.0 (\textpm{} 6.92) \\
sSIFT BoW$_{100}$ & 92.4 (\textpm{} 3.36) & 92.2 (\textpm{} 3.39) & 66.2 (\textpm{} 5.67) & 50.3 (\textpm{} 9.13) \\
dSIFT VLAD$_{32}$ & 92.5 (\textpm{} 3.01) & 94.9 (\textpm{} 2.85) & 74.6 (\textpm{} 5.25) & 74.4 (\textpm{} 9.04) \\
dSIFT VLAD$_{100}$ & 94.0 (\textpm{} 2.55) & 95.9 (\textpm{} 2.59) & 81.2 (\textpm{} 8.0) & 76.9 (\textpm{} 11.3) \\
sSIFT VLAD$_{32}$ & 95.3 (\textpm{} 2.46) & 96.1 (\textpm{} 2.31) & 80.1 (\textpm{} 4.03) & 75.4 (\textpm{} 12.1) \\
sSIFT VLAD$_{100}$ & 95.7 (\textpm{} 2.31) & 96.8 (\textpm{} 2.42) & 84.1 (\textpm{} 5.36) & 79.0 (\textpm{} 8.95) \\
VGG$_{4}$ VLAD$_{64}$ & -- & 97.9 (\textpm{} 1.88) & -- & 84.6 (\textpm{} 8.72) \\
VGG$_{4}$ VLAD$_{32}$ & 96.6 (\textpm{} 2.31) & 98.1 (\textpm{} 1.78) & 88.7 (\textpm{} 3.32) & 84.1 (\textpm{} 6.14) \\
VGG$_{5}$ VLAD$_{32}$ & 94.7 (\textpm{} 2.82) & 98.5 (\textpm{} 1.46) & 76.4 (\textpm{} 5.41) & 78.5 (\textpm{} 8.82) \\
VGG$_{5}$ VLAD$_{64}$ & -- & 98.9 (\textpm{} 1.17) & -- & 83.1 (\textpm{} 6.63) \\
mVGG$_{4}$ VLAD$_{32}$ & -- & 98.2 (\textpm{} 1.71) & -- & 81.0 (\textpm{} 9.58) \\
mVGG$_{5}$ VLAD$_{32}$ & -- & 98.3 (\textpm{} 1.57) & -- & 80.5 (\textpm{} 8.66) \\
\end{tabular}
\end{table*}

Using raw (normalized) flattened images as feature vectors affords around 50\% accuracy on the microconstituent classification task -- substantially better than the expected performance of a random classifier (33\%).
The baseline s-SIFT BoW method with $\chi^2$ kernel attains a solid 90\% accuracy on the microconstituent classification task, somewhat higher than the accuracy of $83 \pm 3\%$ recently reported for the same method applied to a much smaller (but more diverse) analogous 7-class microstructure classification task\cite{decost2015}.
Switching to dense SIFT features slightly decreases the classification accuracy, especially for the cropped image set and with a smaller dictionary size.

VLAD-encoding improves the average SIFT-based image representation performance by up to an additional 6\% for the microconstituent classification task, compared to the $\chi^2$ BoW method.
In the future, it may be interesting to explore competitive alternative normalization schemes for VLAD and Fisher encoding (i.e. the power normalization method\cite{perronnin2010}, and applying the Hellinger kernel and PCA to individual SIFT features before the dictionary encoding step\cite{ke2004,arandjelovic2012,chatfield2011}).

The CNN-derived features we investigated consistently offer the best classification performance, though the marginal improvement of over VLAD-encoded SIFT features is roughly equal to the sample standard deviation.
The marginal gain in classification performance results from moving from block4 features to higher-level CNN features is slight.

The performance differences between methods are much greater for the more difficult task of annealing condition classification.
The variance in the performance estimate for any given method is also much higher for this task, primarily due to the very small dataset size.
Again, the raw images yield classification accuracies that are substantially greater than the expected performance of a random classifier ($1/13 \simeq 7.7\%$).
The SIFT-based BoW representations outperform the raw images by a much wider margin than for the microconstituent task.
VLAD-encoding the SIFT features effectively doubles the accuracy compared to BoW encoding with the same dictionary size.
It's clear from these results that the higher-level CNN features yield abstract microstructure representations that better capture to variations in annealing and imaging conditions, as further discussed in \ref{sec:datavis}.

It is interesting to note that sparse SIFT representations yield classification accuracies as high or higher than dense SIFT representations.
Speculatively, any potential benefit from a higher sampling density of local image features may be countered by the reduced rotation invariance of the fixed-orientation d-SIFT descriptors.
For general microstructure characterization tasks, this rotation invariance is desirable, as any special orientation relationships (such as e.g. the preferred growth directions of Widmanstätten lath) are not necessarily related to the image reference frame.

Perhaps surprisingly, VGG$_{\text{4}}$ features seem to provide more discriminative representations for the spheroidite microstructures than the higher-level VGG$_{\text{5}}$ representations.
Additionally, pooling multiscale VGG features seems plays no significant role on this task.
One possible explanation is that the high-level convolution filters in the CNN are heavily optimized to perform an object recognition task, and provide discriminative abstract representations of objects in natural images--the fully-connected layers encode information about the global geometry of the objects detected in the image\cite{cimpoi2015}.
This can be mitigated by fine-tuning (re-training) the high-level CNN layers\cite{girshick2014}, or by employing additional feature pooling and encoding steps\cite{gong2014,razavian2014,he2015,cimpoi2015}, as we have attempted with VLAD encoding.
An alternative possibility is that the present 3-class UHCS dataset is simply not a challenging classification task relative to current challenges in natural image recognition.
Finally, variation in the labeling of the micrographs could also set an upper limit for classification accuracy on this dataset.

\subsection{Data visualization \label{sec:datavis}}
\label{sec-3-2}
We used t-SNE to visualize the distributions of high-dimensional microstructure features obtained with each feature extraction method.\footnote{Because of space constraints, t-SNE maps and properties plots for many of the image representations investigated in this report were omitted in the manuscript. They are available in the supplementary materials and/or upon request.}
t-SNE is an unsupervised technique; the only input is the set of microstructure representations.
Metadata such as primary microconstituent labels and processing metadata play no role on the structure of the resulting t-SNE maps.
In Sections \ref{sec:tsne-maps} and \ref{sec:tsne-processing} we explore in more detail the image representation yielding the best classification results (VLAD-encoded VGG$_{5}$ features).
Section \ref{sec:tsne-maps} examines local variations in microstructure within the t-SNE representation, and Section \ref{sec:tsne-processing} examines the relationship between microstructure features and the available processing metadata.

Bear in mind that t-SNE explicitly aims to reveal local structure within high-dimensional data, and that distances in the high-dimensional feature space can not always be exactly preserved in the low-dimensional t-SNE map\cite{vandermaaten2008}.
As a result, large distances in the low-dimensional representation do not necessarily indicate large distances in the high-dimensional data, and high-distance `seams' in the low-dimensional representation may exist where it is not possible to cleanly project the high-dimensional data.
However, small distances in the t-SNE representation should correspond to a high degree of visual similarity.

\subsubsection{t-SNE microstructure maps \label{sec:tsne-maps}}
\label{sec-3-2-1}
\begin{figure}[!htbp]
  \centering
  \includegraphics[width=0.45\textwidth]{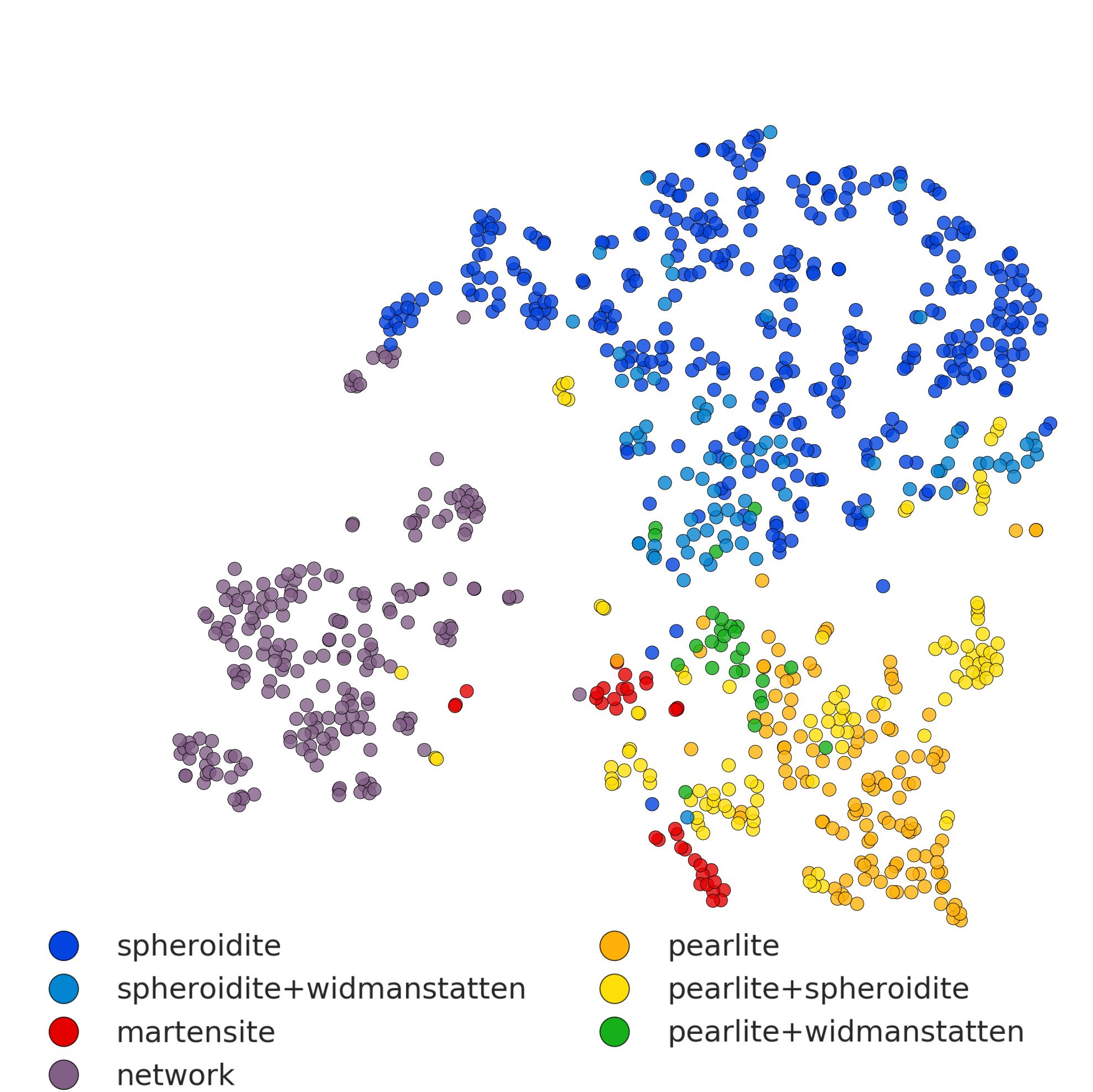} \\
  \caption{The $VGG_5$ $VLAD_{32}$ t-SNE map annotated with primary microconstituent.}
  \label{fig:tsne-microconstituent}
\end{figure}

In order to better understand how the microstructural representations are clustered and related in high dimensional space, we map them in two dimensions via the t-SNE visualization method.
Figure \ref{fig:tsne-microconstituent} shows the t-SNE map for the entire UHCS dataset of 961 full-size images with markers color coded by primary microstructural constituent; the image representation is the VGG-block5 encoding, which was found to have the highest accuracy in the classification task. 
As shown in Figure \ref{fig:tsne-microconstituent}, the spheroidite-related, pearlite-related, and network micrographs each form distinct, extended clusters, with subclusters denoting more closely related images, as discussed below.
The martensitic images form two separate clusters that are closely related to the pearlite structures, as might be expected due to their similar aligned morphologies.The pearlite+widmanstatten images are more loosely clustered among the other pearlite-related images.
A few outlier points in each category indicate micrographs that may challenge the image representation scheme, or may result from noise in the manual primary microconstituent process.

\begin{figure*}[!htbp]
  \centering
  \begin{subfigure}[]{0.6\textwidth}
  \fbox{
  \includegraphics[width=0.98\textwidth,clip=true]{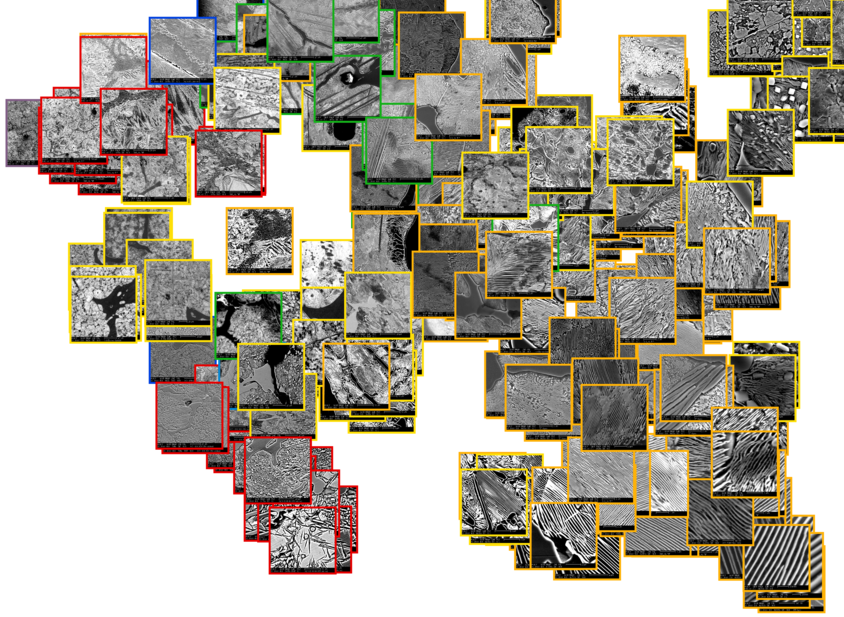}}
  \caption{}
  \label{fig:tsnepearliteim}
  \end{subfigure}
  \begin{subfigure}[]{0.3\textwidth}
  \includegraphics[width=\textwidth,clip=true]{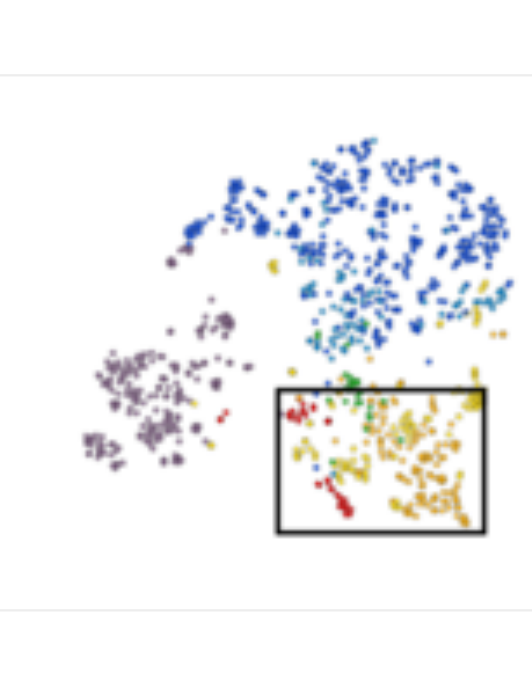}
  \caption{}
  \label{fig:tsnepearliteleg}
  \end{subfigure}
  \caption{(\subref{fig:tsnepearliteim}) VGG-block5 t-SNE microstructure map excerpt showing micrographs containing pearlite (\subref{fig:tsnepearliteleg}) in the lower-right portion of the VGG-block5 t-SNE map with markers coded by primary microconstituent (Figure \ref{fig:tsne-microconstituent}).}
  \label{fig:tsnepearlite}
\end{figure*}

\begin{figure*}[!htbp]
  \centering
  \begin{subfigure}[]{0.6\textwidth}
  \fbox{
  \includegraphics[width=0.98\textwidth]{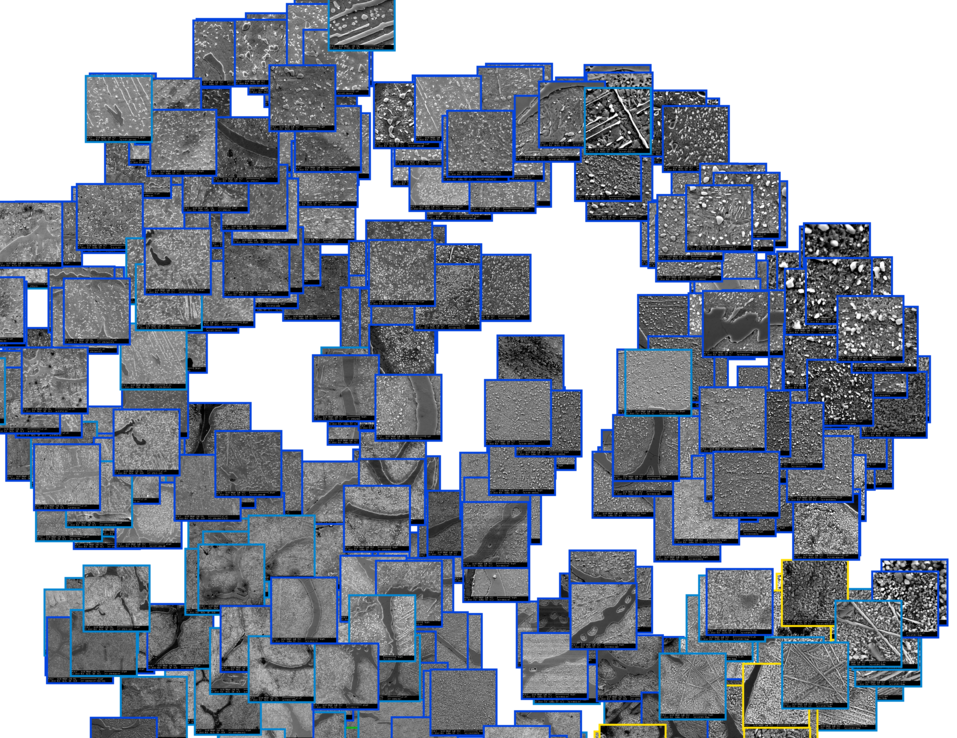}}
  \caption{}
  \label{fig:spheroiditemap}
  \end{subfigure}
  \begin{subfigure}[]{0.3\textwidth}
  \includegraphics[width=\textwidth]{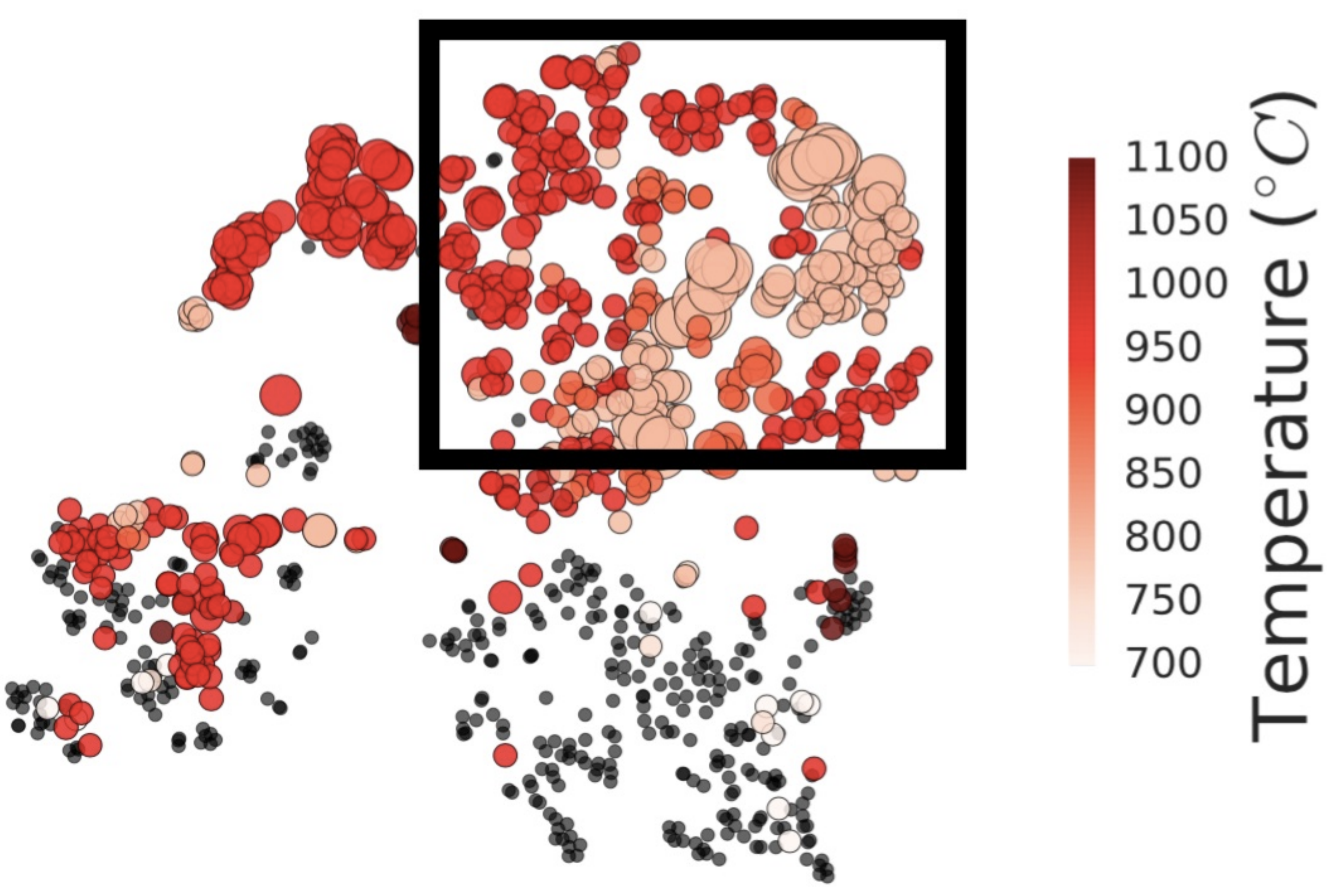}
  \caption{}
  \label{fig:spheroiditeleg}
  \end{subfigure}
  \caption{(\subref{fig:spheroiditemap}) VGG-block5 t-SNE microstructure map excerpt showing micrographs containing spheroidized cementite of various morphologies (\subref{fig:spheroiditeleg}) in the left-most portion of the VGG-block5 t-SNE map, with marker colors indicating annealing temperature and relative marker sizes indicating annealing time (Figure \ref{fig:tsne-tt}).}
  \label{fig:tsnespheroidite}
\end{figure*}

\begin{figure*}[!htbp]
  \centering
  \begin{subfigure}[]{0.6\textwidth}
  \fbox{
  \includegraphics[width=0.98\textwidth]{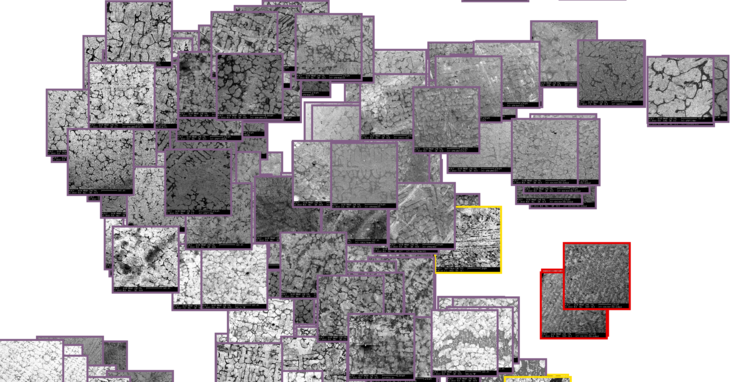}}
  \caption{}
  \label{fig:tsnecarbidenetworkim}
  \end{subfigure}
  \begin{subfigure}[]{0.3\textwidth}
  \includegraphics[width=\textwidth]{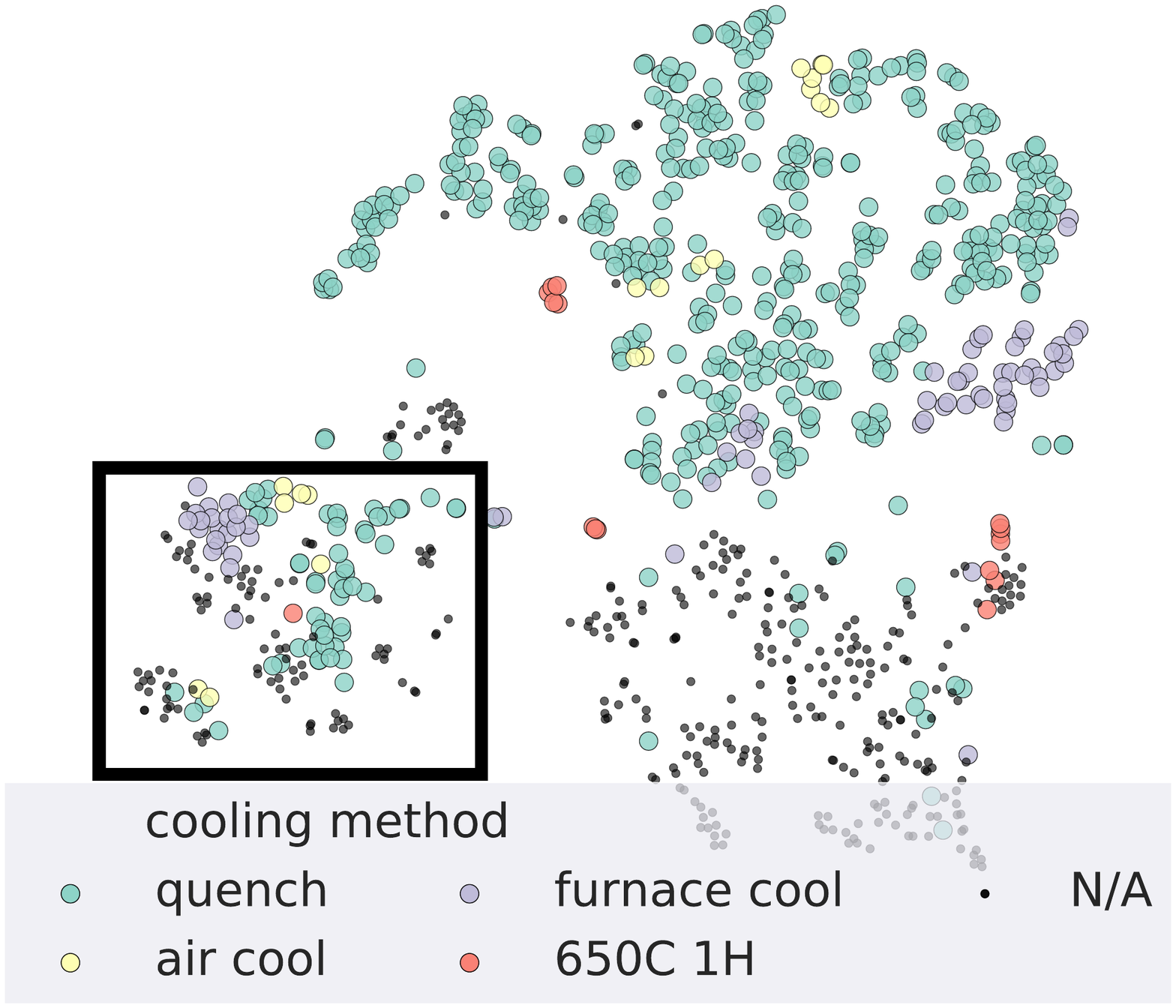}
  \caption{}
  \label{fig:tsnecarbidenetworkleg}
  \end{subfigure}
  \caption{(\subref{fig:tsnecarbidenetworkim}) VGG-block5 t-SNE microstructure map excerpt showing micrographs containing high-level views of the proeutectoid cementite network (\subref{fig:tsnecarbidenetworkleg}) in the lower-right quadrant of the VGG-block5 t-SNE map with markers colored by cooling method (Figure \ref{fig:tsne-cool}).}
  \label{fig:tsnecarbidenetwork}
\end{figure*}

Figures \ref{fig:tsnepearlite}, \ref{fig:tsnespheroidite}, and \ref{fig:tsnecarbidenetwork} show detailed microstructure maps obtained by displaying each micrograph centered on its corresponding position in the VGG-block5 t-SNE map shown in Figure \ref{fig:tsne-microconstituent}.\footnote{The full microstructure map is much too large to display in format of the present paper, but is available in the supplemental materials, along with maps for other microstructure representations.}
The black frames on the inset t-SNE scatter plots indicate which portion of the map is displayed.
Colored frames around each thumbnail image indicate the primary microconstituent label, following the same color map as used in Figure \ref{fig:tsne-microconstituent}.

Figure \ref{fig:tsnepearliteim} focuses on the visual appearance of the high-magnification pearlite micrographs, as indicated by in Figure \ref{fig:tsnepearliteleg}, as well as two apparent clusters of martensite structures at the left.
These pearlite micrographs span multiple orientations, magnification, and lamellar spacing, increasing in complexity from the lower right corner of the map.
The pearlite micrographs in the lower right corner of the map are high-magnification views of individual pearlite domains; traversing up to the top of the figure widens the field of view with more morphological variation, with clear trend in the lamellar orientation.
The pearlite matrix often contains spheroidite in the micrographs in the top half of this figure.

\fboxrule=2pt
\fboxsep=0pt

Figure \ref{fig:spheroiditemap} shows most of the spheroidite micrographs in the block5 t-SNE map.
These micrographs tend to cluster together with other micrographs from the same sample, i.e. the same processing conditions, as shown in Figure \ref{fig:spheroiditeleg}.
The magnification of these micrographs generally increases from the lower-left to the upper right quadrant of this map as well.
Micrographs near at bottom of this map focus on the spheroidite-free denuded zones adjacent to the carbide network, while the higher-magnification micrographs at the top focus on the morphology and spatial distribution of individual spherodite particles.
The microstructures in the upper left portion of this map are the result of higher annealing temperatures compared to the micrographs in the lower right quadrant (see Figure \ref{fig:spheroiditeleg}); this temperature gradient corresponds to a gradient in the spheroidite morphology across the map.
One notable exception to this trend is the cluster of high-temperature microstructures in the lower-right corner of Figure \ref{fig:spheroiditemap}, which contain prominent Widmanstätten lath.
While most of the material in this dataset was cooled by quenching, this cluster of micrographs come from material that was furnace-cooled.

Figure \ref{fig:tsnecarbidenetworkim} shows one the main cluster of proeutectoid cementite network microstructures.
Many of these micrographs were subjected to similar processing conditions, with the bulk of them coming from samples annealed at $970 ^{\circ}C$ for 90 minutes before being either air-cooled or water-quenched (see Figure \ref{fig:tsnecarbidenetworkleg}).
The network structures formed under these common annealing conditions form two distinct groups in the VGG-block5 t-SNE map, in the upper left and central regions of Figure \ref{fig:tsnecarbidenetwork}.
The group at the upper left were furnace-cooled, while the central group were quenched.
The quenched network micrographs clearly have lower contrast than the furnace-cooled network micrographs, and the cementite in the pearlitic matrix has a somewhat different morphology.

With its ability to sensibly arrange high-dimensional image representations in a two-dimensional map, t-SNE is a valuable visualization method for microstructural image datasets, and for understanding the efficacy of high-dimensional microstructure representations.
It enables quick visual scans to identify related images at large and small scales, captures systematic trends in microstructural morphology, and when coupled with processing metadata can graphically display the microstructure - processing link, as we discuss further in the next section.

\subsubsection{Processing metadata \label{sec:tsne-processing}}
\label{sec-3-2-2}
Though a regression model relating microstructural outcomes back to processing variables would be a more relevant model for a microstructure design task, the present dataset has an unbalanced distribution of processing parameters.
However, examining these processing parameters by microstructural category still yields quantitative insight into the ability of the computer vision approach to infer processing - microstructure relationships.
To this end, we map processing metadata onto the t-SNE map for image representations and explore the systematic trends between structure and processing.

\begin{figure*}[!htbp]
  \centering
  \hfill
  \begin{subfigure}[]{0.3\textwidth}
  \includegraphics[width=\textwidth]{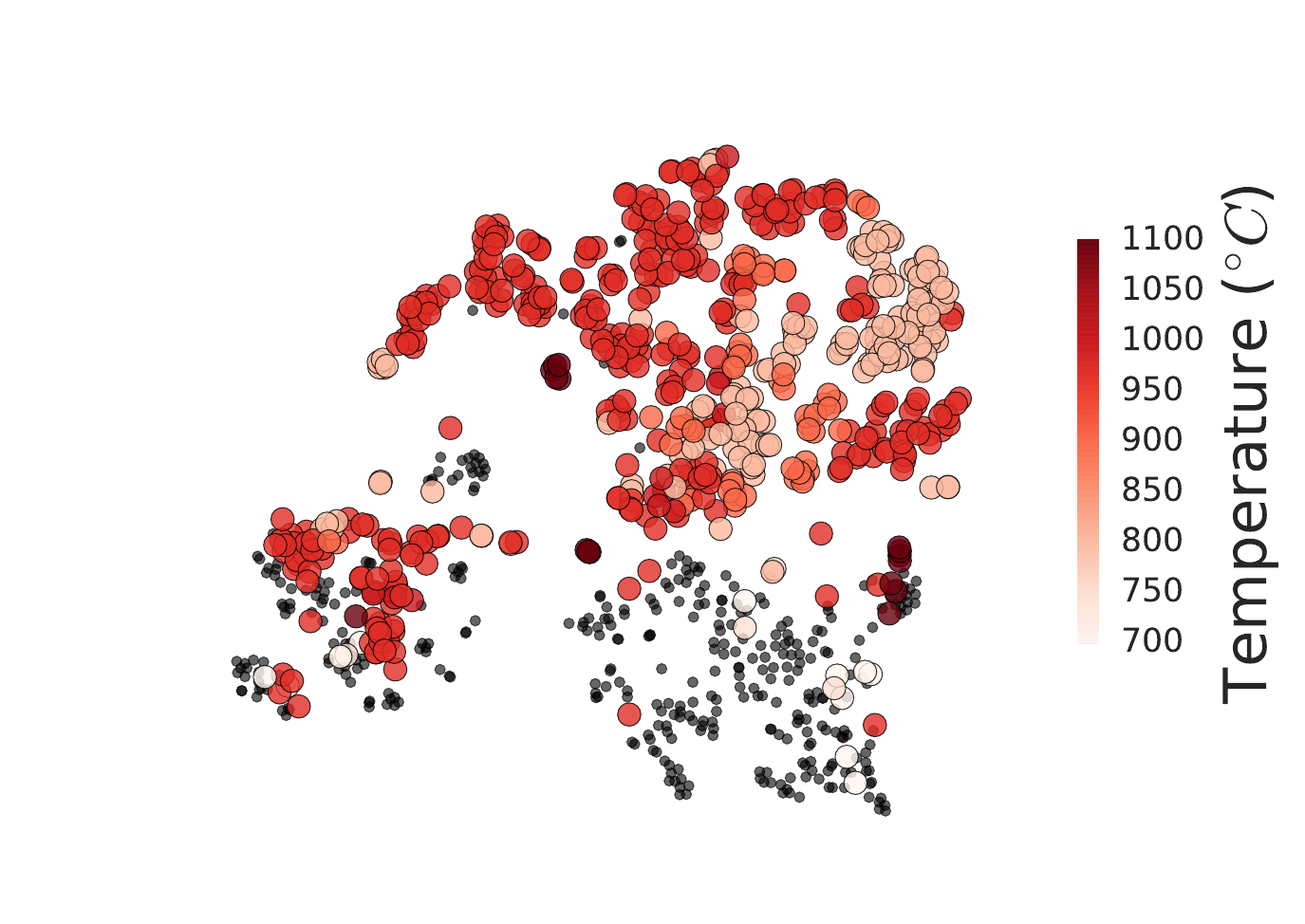}
  \caption{}
  \label{fig:tsne-temp}
  \end{subfigure} \hfill
  \begin{subfigure}[]{0.3\textwidth}
  \includegraphics[width=\textwidth]{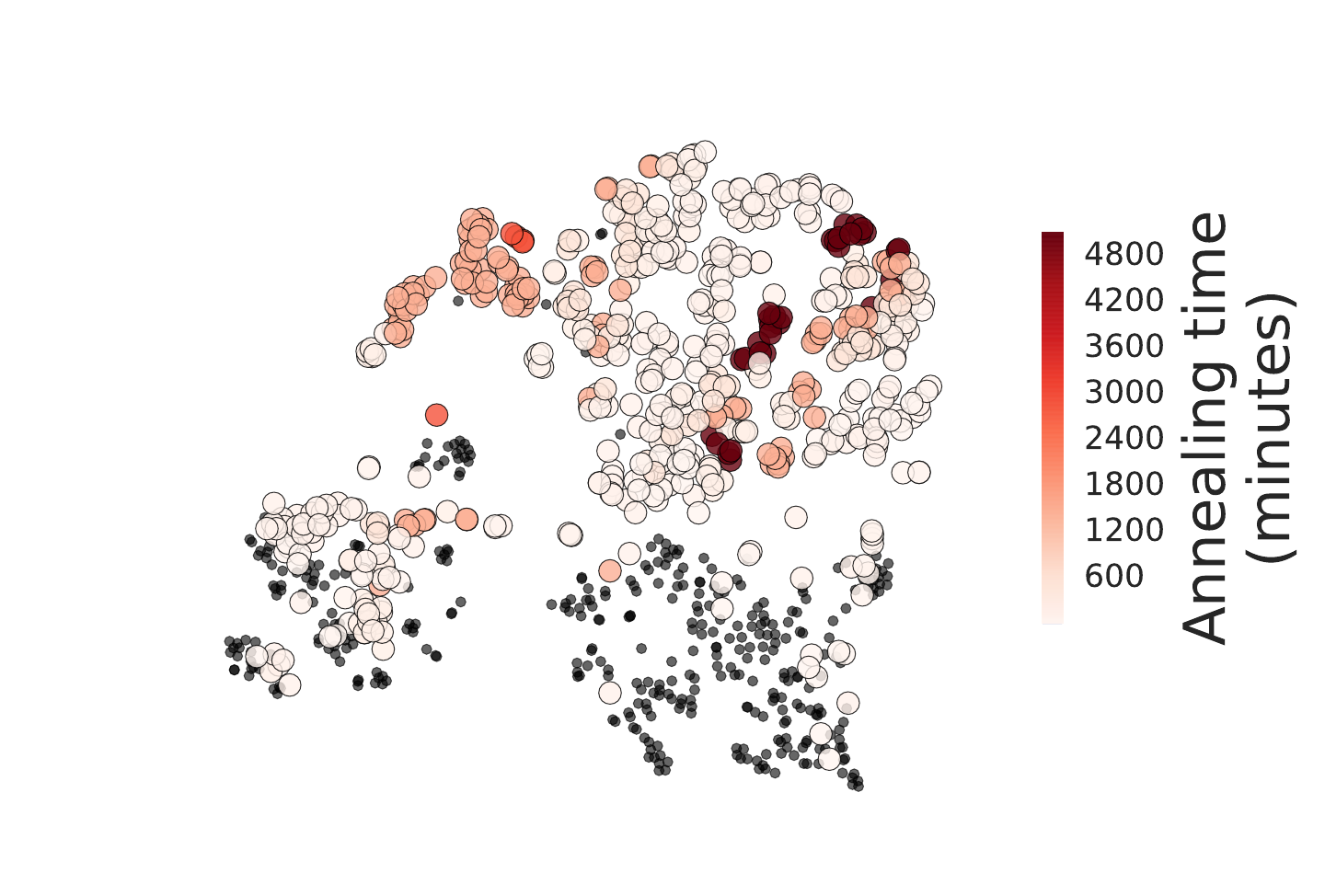}
  \caption{}
  \label{fig:tsne-time}
  \end{subfigure} \hfill
  \begin{subfigure}[]{0.3\textwidth}
  \includegraphics[width=\textwidth]{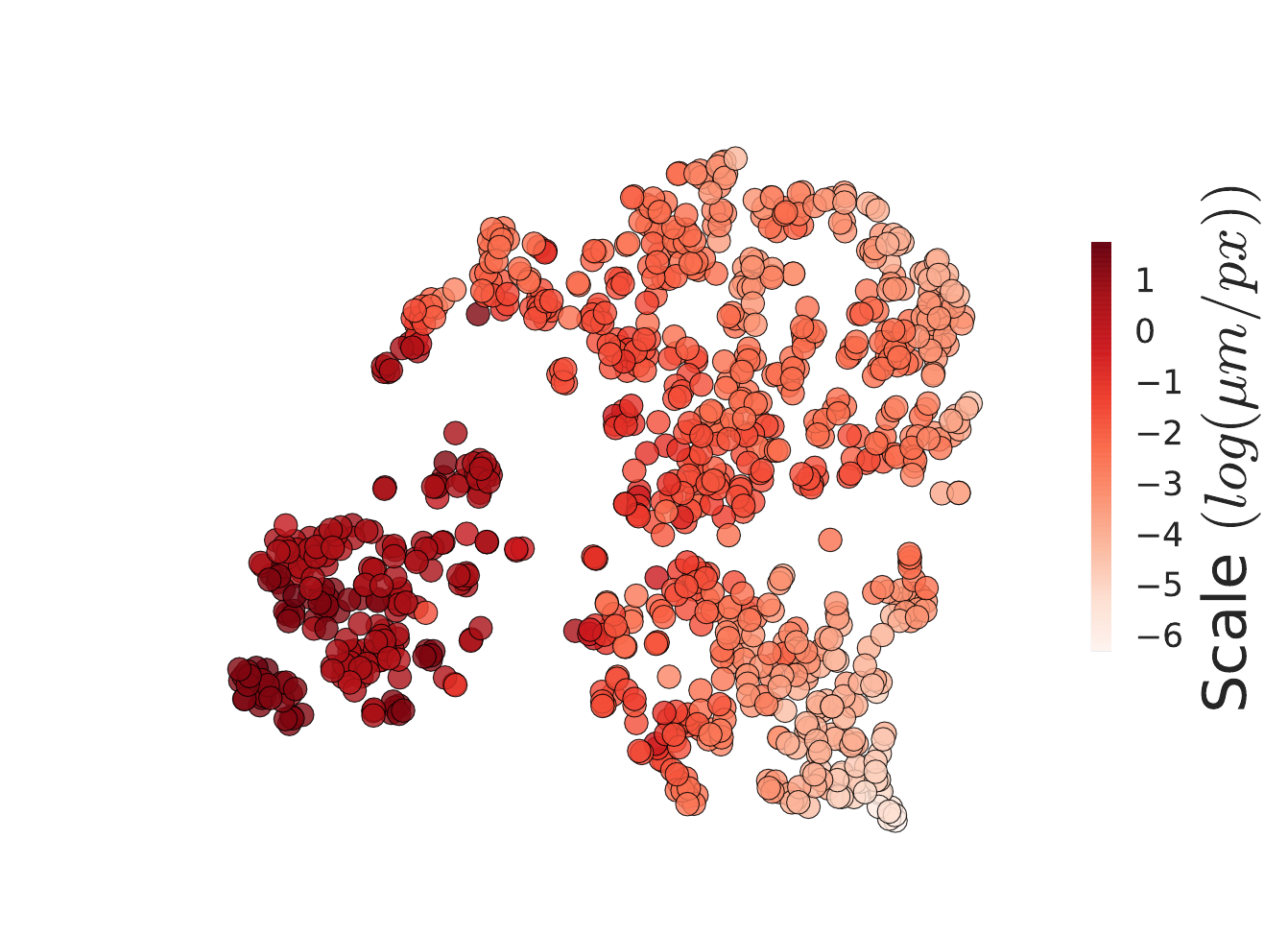}
  \caption{}
  \label{fig:tsne-scale}
  \end{subfigure} \hfill \\ \hfill
  \begin{subfigure}[]{0.4\textwidth}
  \includegraphics[width=\textwidth]{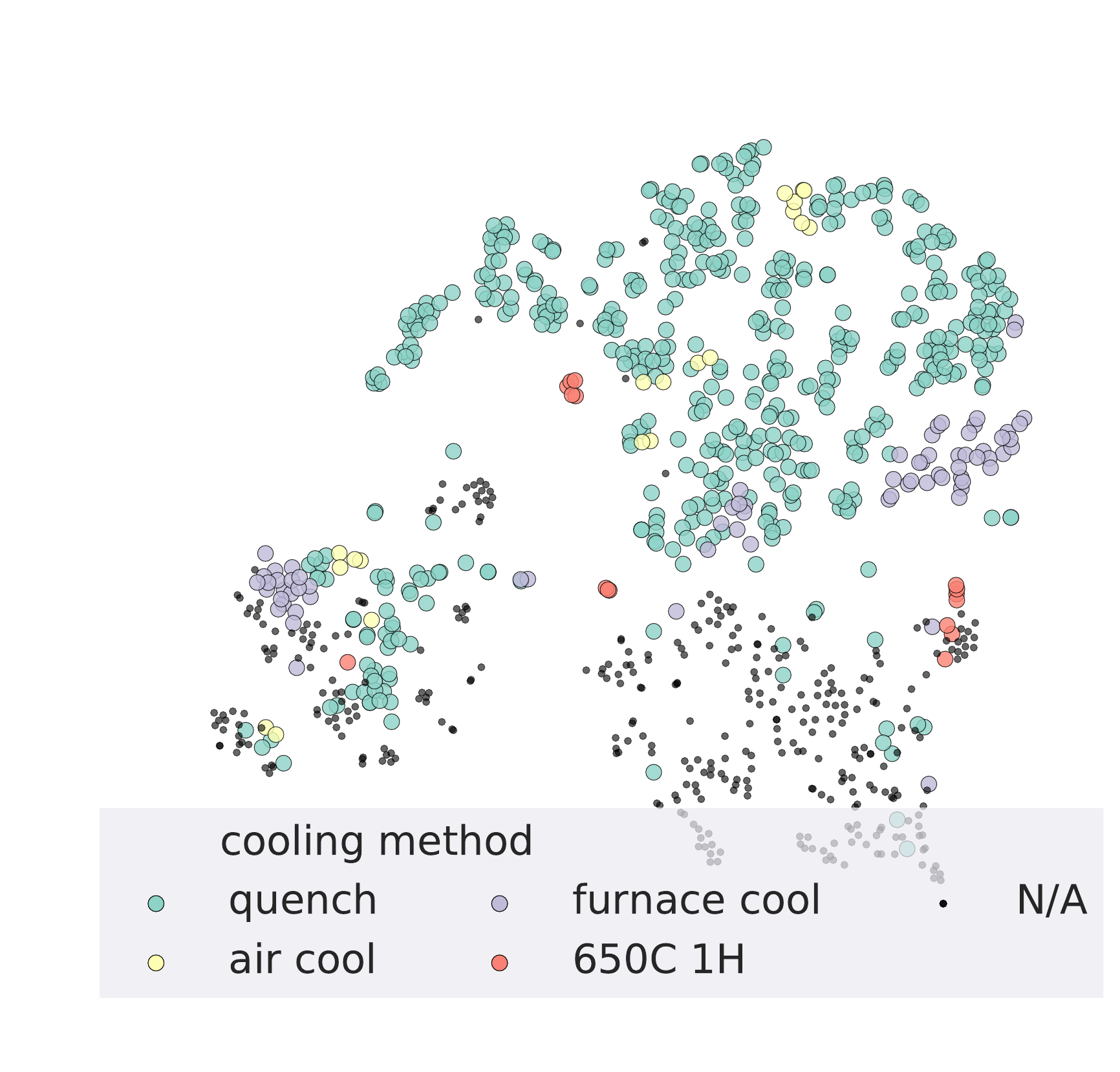}
  \caption{}
  \label{fig:tsne-cool}
  \end{subfigure} \hfill
  \begin{subfigure}[]{0.4\textwidth}
  \includegraphics[width=\textwidth]{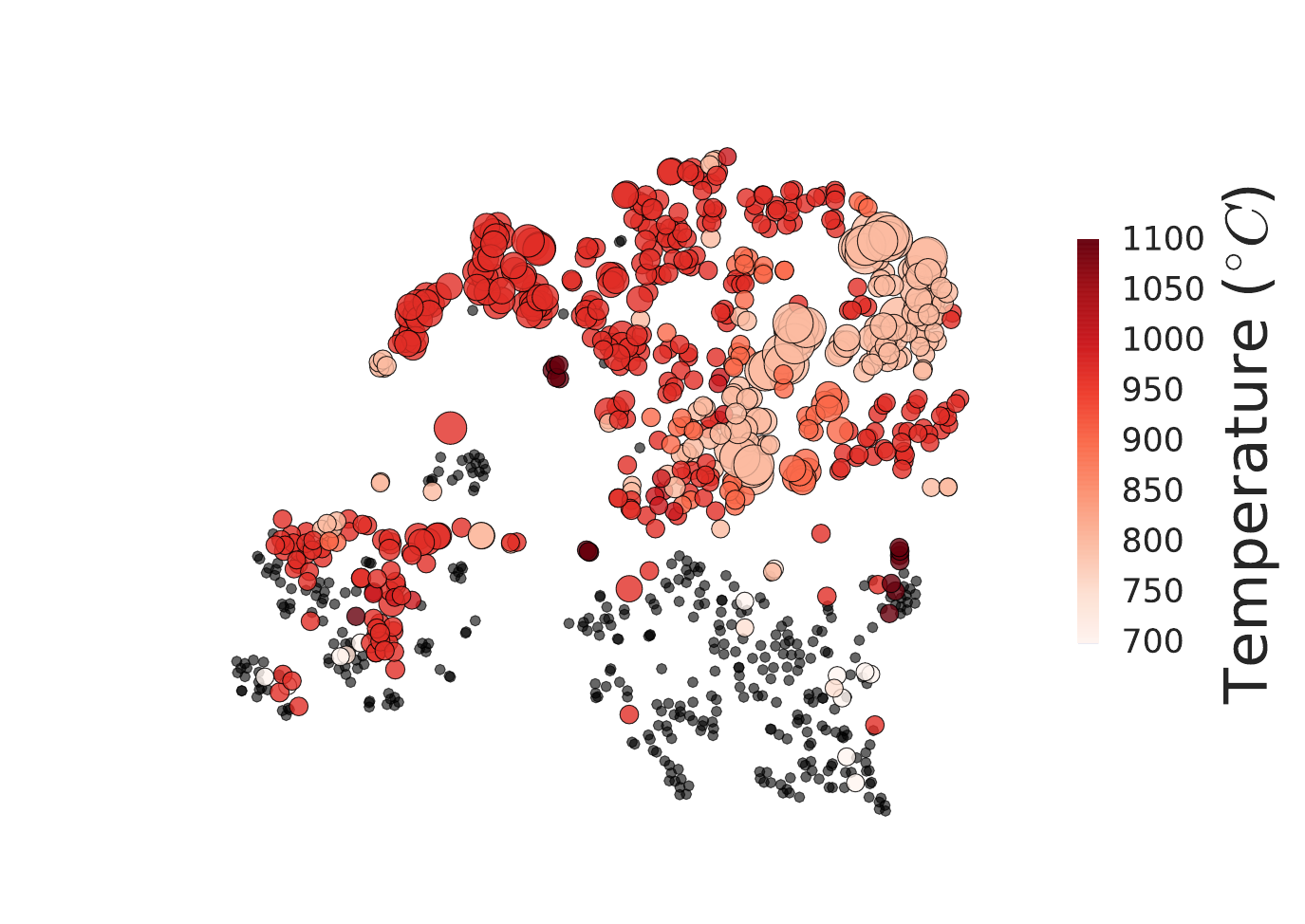}
  \caption{}
  \label{fig:tsne-tt}
  \end{subfigure} \hfill
  \caption{The $VGG_5$ $VLAD_{32}$ t-SNE map from Figure \ref{fig:tsne-microconstituent} annotated with processing metadata. Best viewed electronically. (\subref{fig:tsne-temp}) Temperature (\subref{fig:tsne-time}) annealing time, (\subref{fig:tsne-scale}) magnification in $\log_{10} (\mu m / px)$, (\subref{fig:tsne-cool}) quench method, and (\subref{fig:tsne-tt}) a bubble plot with colors indicating annealing temperature and marker size proportional to annealing time. For reference, the micrograph resolutions range from $0.002 \mu m / px$ to $5.9 \mu m / px$.}
  \label{fig:processing}
\end{figure*}

Figure \ref{fig:processing} illustrates the relationships between the available annealing schedule metadata and the resulting microstructure as shown by the VGG$_{\text{5}}$ t-SNE map from Figure \ref{fig:tsne-microconstituent}.
Figure \ref{fig:tsne-temp} shows the annealing temperature in $^{\circ}C$; Figure \ref{fig:tsne-time} shows the annealing time in minutes; Figure \ref{fig:tsne-scale} shows the magnification (in microns per pixel) on a logarithmic scale; Figure \ref{fig:tsne-cool} shows the cooling rate; and Figure \ref{fig:tsne-tt} jointly illustrates the annealing time (proportional to marker size) and temperature (indicated by the color map).
The small black markers indicate micrographs for which no processing metadata is currently available; these are mostly the high-magnification pearlite matrix micrographs, along with a subset of the network micrographs.

Apparent cluster structure in the VGG$_{\text{5}}$ t-SNE map clearly relates qualitatively to the annealing time and temperature data shown in Figure \ref{fig:tsne-tt}.
Most of the tightest local clusters in the t-SNE map consist of microstructures with the same or very similar annealing schedules.
However, the magnification also plays a significant role.
Consider the three small clusters of large, cream-colored points in the upper right quadrant of Figure \ref{fig:tsne-tt} (low-temperature, long anneal micrographs of spheroidite) and their corresponding points in Figure \ref{fig:tsne-scale}.
The processing parameters and microstructures are similar between all three clusters, but the micrograph magnification increases by a factor of two moving from the upper right cluster to the middle cluster, and by yet another factor of two moving to the lower left cluster.
The s-SIFT BoW and  VLAD representation both also display this same effect; interestingly the VGG-pool5 representation seems to be somewhat more robust to changes in magnification, even though the scale-invariance of this method should be weaker and more implicit.
Pooling the VGG$_{\text{5}}$ feature maps over multiple scales improves the situation (see the supplemental materials) by bringing the corresponding images to the upper two clusters closer together; the third cluster is still quite distinct, as those micrographs include substantially wider fields of view focusing on the carbide network and the morphology of the surrounding spheroidite.
Thus, the question of how to incorporate the absolute physical scale of microstructure features into image representations adopted from the object and scene recognition communities must be addressed in order for these methods to help scientists and engineers develop quantitative processing--structure--properties mappings.

\section{Conclusions}
\label{sec-4}
In this report, we establish a dataset for microstructure informatics that focuses on complex, hierarchical, and technologically-relevant microstructures.
We evaluate applications of multiple image representation techniques from the field of computer vision in conjunction with both supervised and unsupervised microstructure informatics tasks.
For this dataset, we show that appropriately pooled and encoded local features (SIFT) and domain-transferred deep convolutional neural network representations can provide classification accuracy better than 95\%.
We also discuss data visualization techniques (t-SNE) for exploratory analysis of microstructure and processing/properties metadata datasets.
Explicit incorporation of the physical scale of microstructure features may be necessary for more quantitative microstructure science applications.

\section*{Acknowledgements}
We gratefully acknowledge funding for this work through National Science Foundation grants DMR-1307138 and DMR-1501830, and through the John and Claire Bertucci Foundation.
The UHCS micrographs were graciously provided by Matthew Hecht, Yoosuf Picard, and Bryan Webler (CMU).
The open source software projects VLFeat\cite{vlfeat}, Scikit-Learn\cite{sklearn}, keras\cite{keras}, and the reference implementation of t-SNE were essential to this work.

\bibliographystyle{unsrt}
\bibliography{uhcs,comparison}
\end{document}